\documentclass[11pt]{article}
\usepackage{moriond,epsfig}
\usepackage{subfigure} 
\def\Journal#1#2#3#4{{#1} {\bf #2}, #3 (#4)}

\def\PRD{{\em Phys. Rev.} D}

\def\be{\begin{equation}}
\def\ee{\end{equation}}
\def\bea{\begin{eqnarray}}
\def\eea{\end{eqnarray}}

\begin{document}
\vspace*{4cm}
\title{MEASUREMENT OF THE UNDERLYING EVENT AT TEVATRON}

\author{DEEPAK KAR, On behalf of the CDF Collaboration}

\address{IKTP, TU Dresden, Germany}

\maketitle\abstracts{CDF Run II data for the underlying event
associated with Drell-Yan lepton pair production are examined as a function of the
lepton-pair transverse momentum. The data are compared with a previous analysis
on the behavior of the underlying event in high transverse momentum jet production and also 
with several other QCD Monte-Carlo models. The goal is
to provide data that can be used to tune and improve the QCD Monte-Carlo models of the
underlying event, which is especially important now in view of the LHC startup.
}

\section{Introduction: the Underlying Event}

In order to find `new' physics at a hadron-hadron collider it is essential to have Monte-Carlo models that simulate accurately the `ordinary' QCD hard-scattering events. To do this
one must not only have a good model of the hard scattering part of the process, but also of the
underlying event.

A typical $2$-to-$2$ hard scattering event is a proton-antiproton collision at the hadron colliders as shown in the Figure 1(a), all happening inside the radius of a proton. In addition to the two hard scattered outgoing partons, which fragment into jets - there is initial and final state radiation (caused by bremsstrahlung and gluon emission), multiple parton interaction (additional $2$-to-$2$ scattering within the same event), `beam beam remnants' (particles that come from the breakup of the proton and antiproton, from the partons not participating in the primary hard scatter). We define the `underlying event' \cite{UE} as everything except the hard scattered components, which includes the `beam-beam remnants' (or the BBR) plus the multiple parton interaction (or the MPI). However, it is not possible on an event-by-event basis to be certain which particles came from the underlying event and, which particles originated from the hard scattering. The `underlying event' ({\it i.e.} BBR plus MPI) is an unavoidable background to most collider observables.  For example, at the Tevatron both the inclusive jet cross section and the b-jet cross section, as well as isolation cuts and the measurement of missing energy depend sensitively on the underlying event. A good understanding of it will lead to more precise measurements at the Tevatron and the LHC.

For Drell-Yan lepton pair production, we have the outgoing lepton anti-lepton pair in the final state and there would be no colored final state radiation. Hence it provides a very clean way to study the underlying event.

\begin{figure}
     \centering
     \subfigure{
             \includegraphics[width=.6\textwidth]{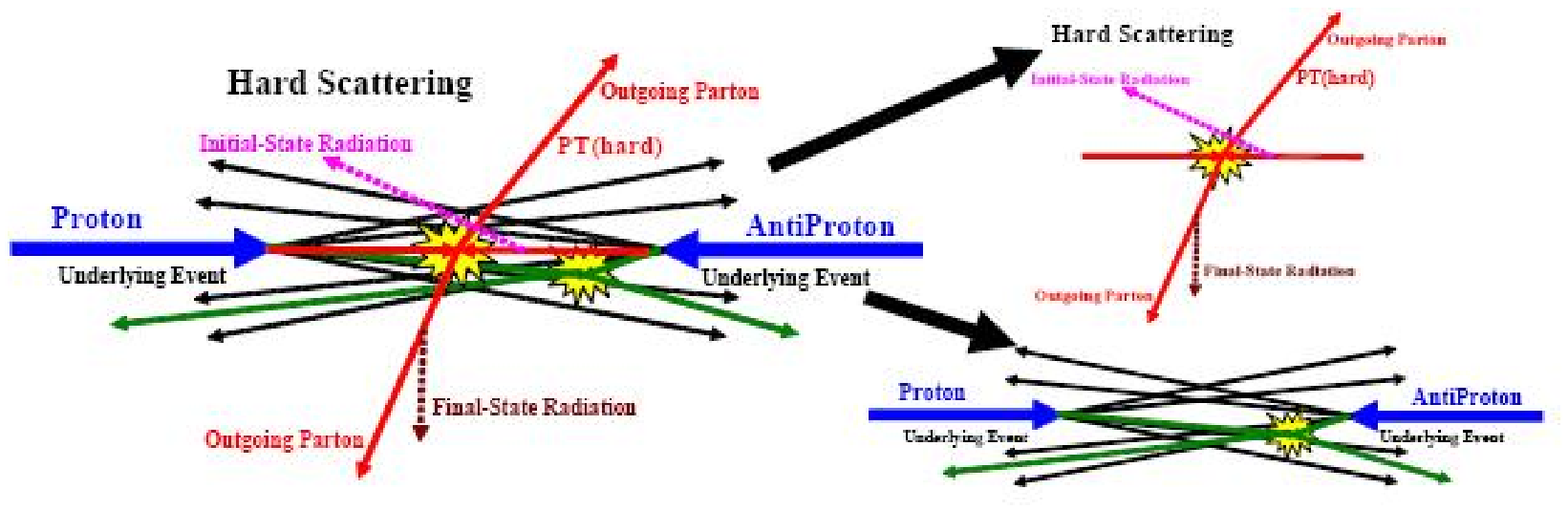}}
     \subfigure{
              \includegraphics[width=.2\textwidth]{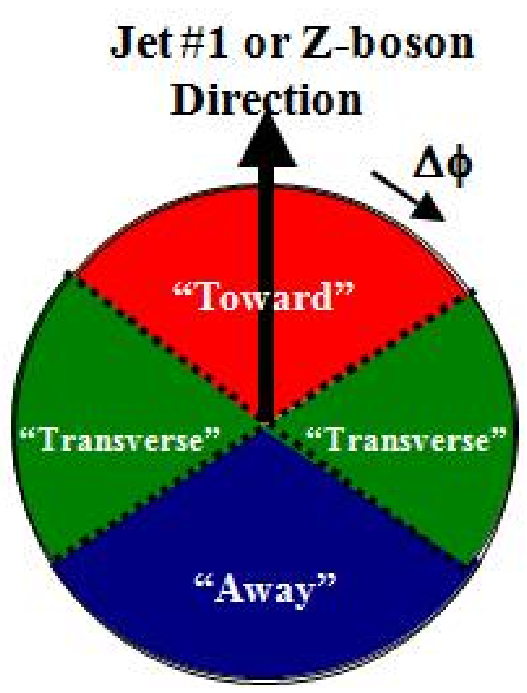}}\\
       \caption{A typical 2-2 hard scattering process and dividing the central region}
\end{figure}

\section{Comparing data with QCD Monte Carlo Models}

\subsection{The Underlying Event as a function of lepton pair $p_T$}

We looked at the charged particles in the range $p_T > 0.5~GeV/c$ and $|\eta| <1$, at the region of Z-boson, defined as $70~GeV/c^2 < M_{ll}<110~GeV/c^2$, in the `toward', `away' and `transverse' regions, as defined in Fig. 1(b). The underlying event observables are found to be reasonably flat with the increasing lepton pair transverse momentum in the transverse and toward regions, but goes up in the away region to balance the lepton pairs.
In Fig. 2(a) and Fig. 2(b), we looked at the two observables corresponding to the underlying event, the number of charged particle density and the charged transverse momentum sum density in the transverse region, compared with {\sc pythia} tunes A and AW \cite{Tunes1}, {\sc herwig} \cite{HERWIG} without MPI and a previous CDF analysis on leading jet underlying event results. We mostly observed very good agreements with {\sc pythia} tune AW Monte Carlo predictions ({\sc herwig} produces much less activity), although the agreement between theory and data is not perfect. We also compared them with leading jet underlying event results and observed reasonably close agreement  - which may indicate the universality of underlying event modeling.

\begin{figure}
     \centering
     \subfigure{
             \includegraphics[width=.3\textwidth, angle =270]{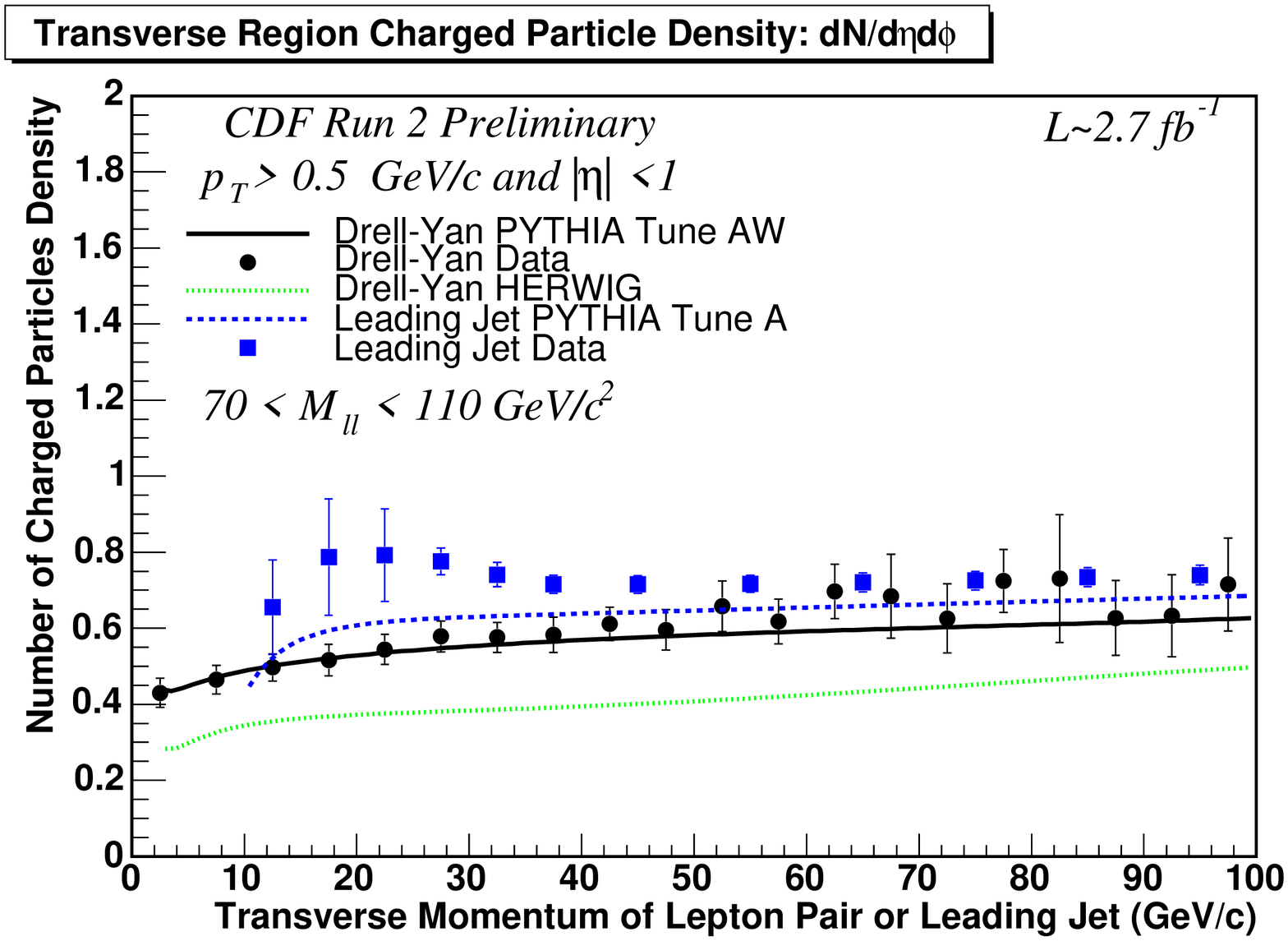}}
     \hspace{.3in}
     \subfigure{
              \includegraphics[width=.3\textwidth, angle=270]{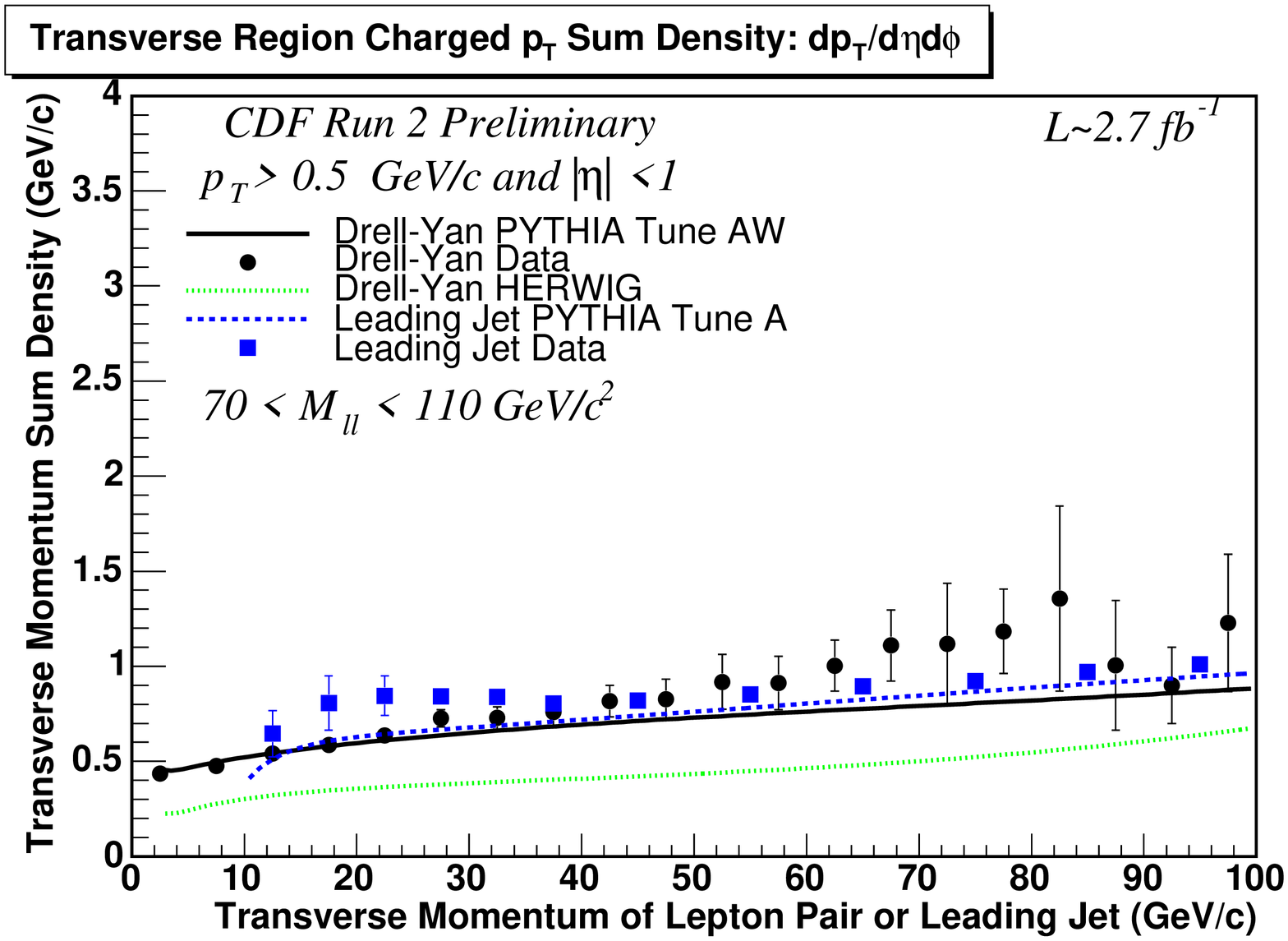}}\\
       \caption{Drell-Yan underlying event plots, charged particle multiplicity on the left and the charged $p_T$ sum on the right}
\end{figure}

\subsection{Correlation Studies}

The rate of change of $<p_T>$ versus charged multiplicity is a measure of the amount of hard versus soft
processes contributing to collisions and it is sensitive to the modeling of the multiple parton
interactions \cite{Corr}. This variable is one of the most sensitive to the combination
of the physical effects present in minimum-bias collisions and is
also the most poorly reproduced variable by the available
Monte Carlo generators. If only the soft beam-beam remnants contributed to
min-bias collisions then $<p_T>$ would not depend on charged multiplicity. If one has two
processes contributing, one soft (beam-beam remnants) and one hard (hard 2-to-2 parton-parton
scattering), then demanding large multiplicity would preferentially select the hard process
and lead to a high $<p_T>$. However, we see that with only these two processes $<p_T>$ increases
much too rapidly as a function of multiplicity. Multiple-parton interactions
provides another mechanism for producing large multiplicities that are harder than the beam-beam
remnants, but not as hard as the primary 2-to-2 hard scattering. 

\begin{figure}
     \centering
     \subfigure{
             \includegraphics[width=.4\textwidth]{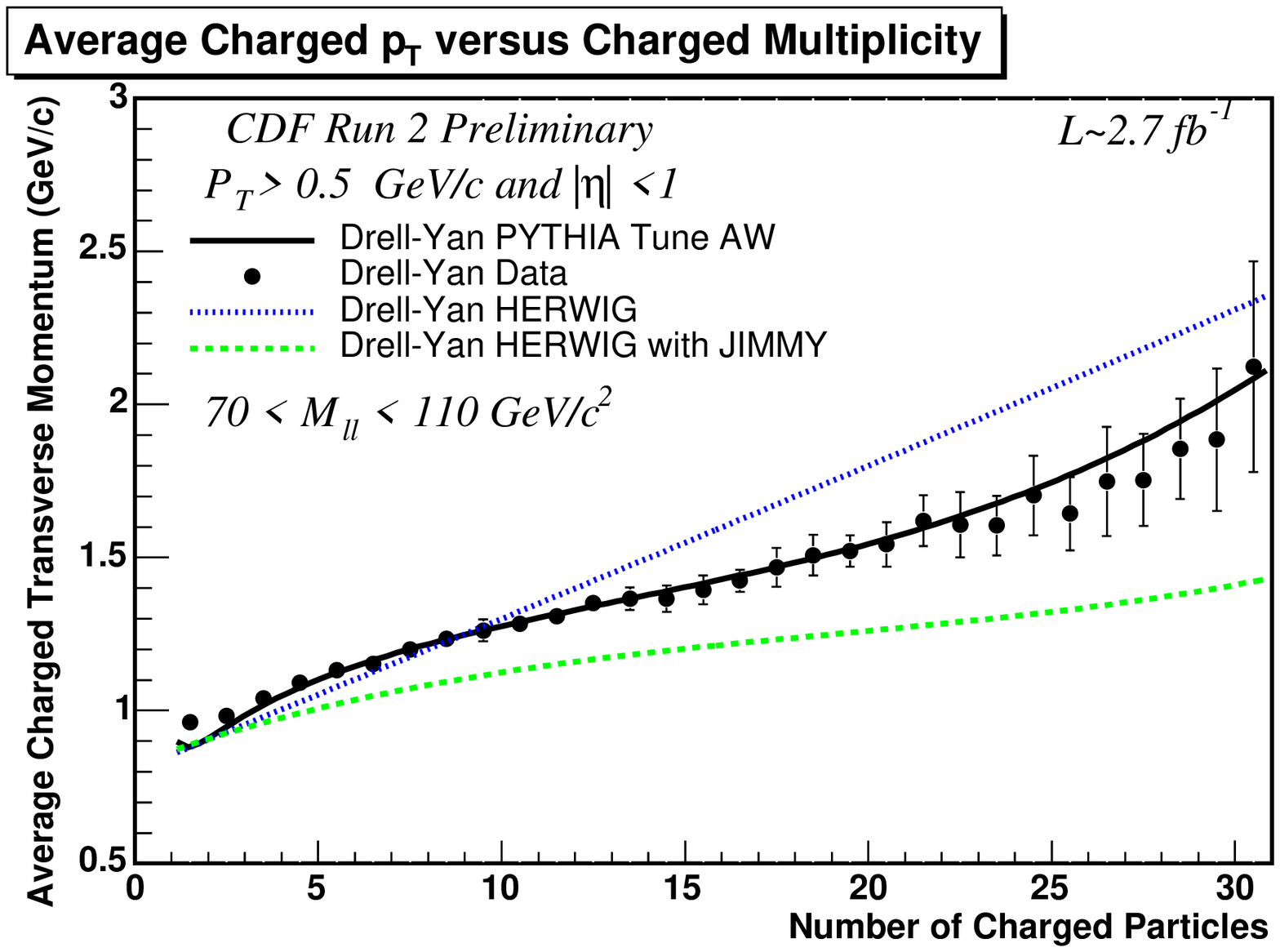}}
     \hspace{.3in}
     \subfigure{
              \includegraphics[width=.4\textwidth]{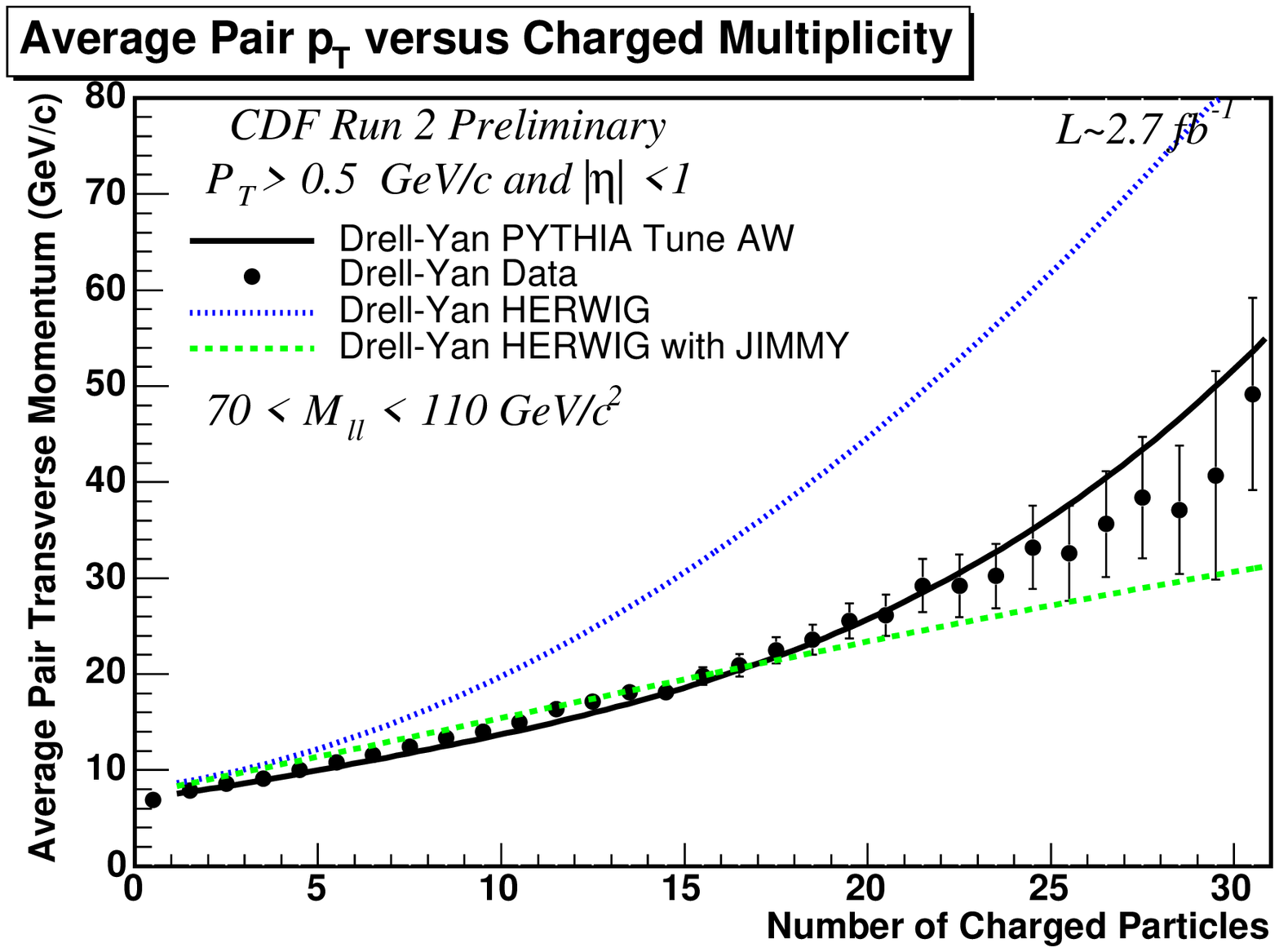}}\\
     \vspace{.2in}
     \subfigure{
                \includegraphics[width=.4\textwidth]{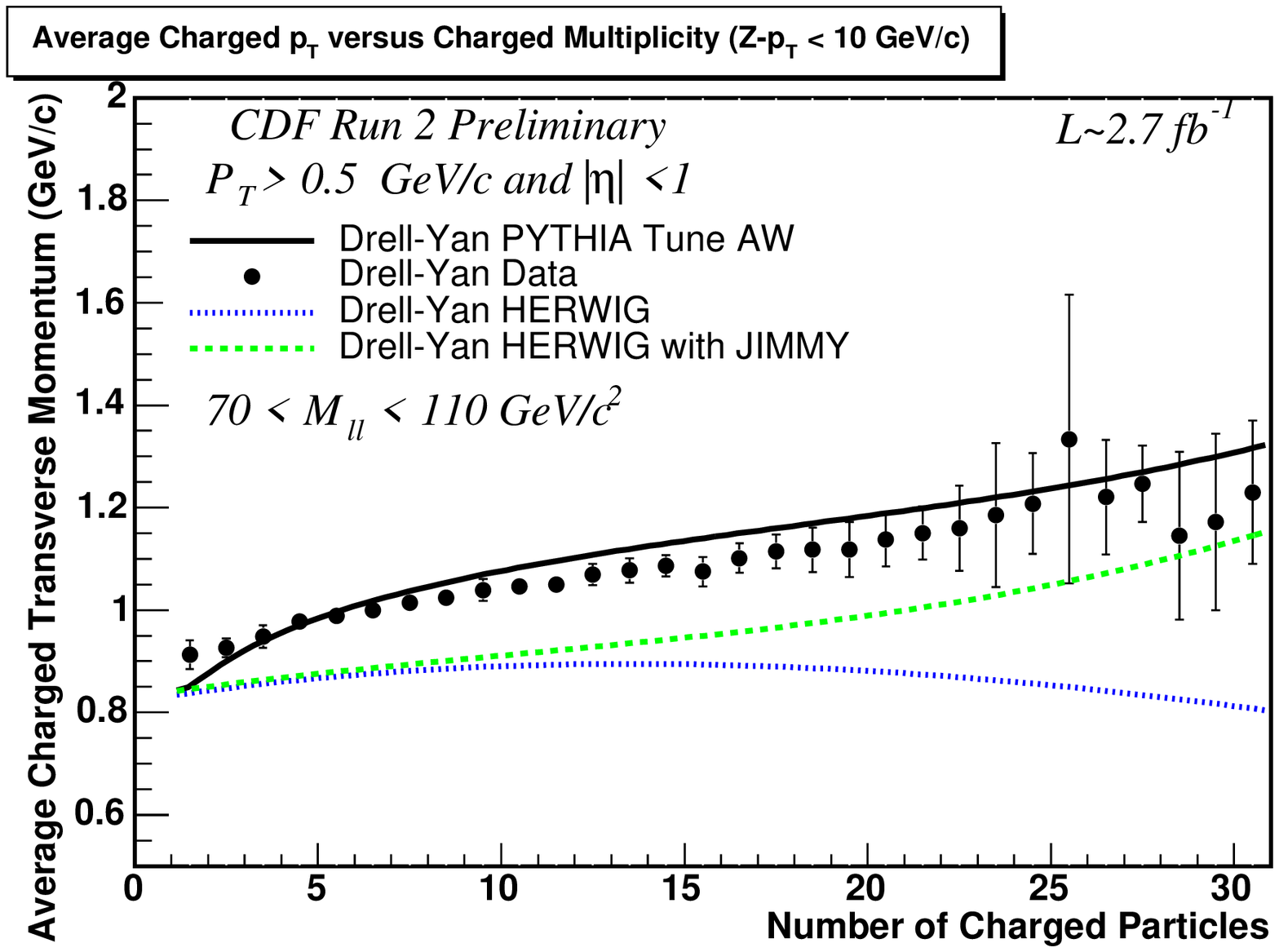}}
     \caption{Charged multiplicity against charged transverse momentum average correlation plots.}
\end{figure}

Fig. 3(a) shows the data corrected to the particle level on the average $p_T$ of
charged particles versus the multiplicity for charged particles with $p_T > 0.5~GeV/c$ and $|\eta| < 1$ for
Z-boson events from this analysis. {\sc herwig} (without MPI) predicts the $<p_T>$ to rise too
rapidly as the multiplicity increases. For {\sc herwig} (without MPI) large multiplicities come from events with a
high $p_T$ Z-boson and hence a large $p_T$ `away-side' jet. This can be seen clearly in Fig. 3(b)
which shows the average $p_T$ of the Z-boson versus the charged multiplicity. Without MPI
the only way of getting large multiplicity is with high $p_T$(Z) events. For the models with MPI
one can get large multiplicity either from high $p_T$(Z) events or from MPI and hence $<p_T(Z)>$
does not rise as sharply with multiplicity in accord with the data. {\sc pythia} tune AW describes
the Z-boson data fairly well.
Fig. 3(c) shows the data corrected to the particle level on the average $p_T$ of
charged particles versus the multiplicity for charged particles with $p_T > 0.5~GeV/c$ and $|\eta| < 1$ for
Z-boson events in which $p_T(Z) < 10 ~GeV/c$. Regardless of all the improvements in the comprehension of low-$p_T$ production, the models are still unable to reproduce second order quantities such as final state particle correlations. We see that $<p_T>$ still increases as the multiplicity increases although not as fast. If we require $p_T(Z) < 10~GeV/c$, then {\sc herwig} (without MPI) predicts that the $<p_T>$ decreases slightly as the multiplicity increases. This is because without MPI and without the high $p_T$ `away-side' jet which is suppressed by requiring low $p_T(Z)$, large
multiplicities come from events with a lot of initial-state radiation and the particles coming from
initial-state radiation are `soft'. {\sc pythia} tune AW describes the behavior of $<p_T>$ versus the multiplicity fairly well even when we select $p_T(Z) < 10~GeV/c$. This strongly suggests that MPI are playing an important role in both these processes.

\section{Summary and Conclusions}

We are making good progress in understanding and modeling the softer physics. CDF tunes A and AW describe the data very well, although we still do not yet have a perfect fit to all the features of the CDF underlying event and min-bias data. Future studies should focus on tuning the energy dependence for the event activity in both minimum bias and the underlying event, which at the moment seems to be one of the least understood aspects of all the models. The underlying event is expected to be much more active in LHC and it is critical to have sensible underlying event models containing our best physical knowledge and intuition, tuned to all relevant available data.

\end{document}